\documentclass[10pt,twocolumn]{article}

\usepackage{multicol}
\usepackage{graphicx}
\usepackage{amsmath}
\usepackage{lipsum}
\usepackage{color}
\usepackage{siunitx}
\usepackage{authblk}
\usepackage[margin=2cm]{geometry}

\title{Self-crumpling elastomers: bending induced by the drying stimulus of a nanoparticle suspension}
\author[1]{Fran\c{c}ois Boulogne}
\author[1]{Howard A. Stone}
\affil[1]{ Department of Mechanical and Aerospace Engineering, Princeton University, Princeton, NJ 08544}
\date{\today}
\begin{document}


\twocolumn[
    \begin{@twocolumnfalse}
        \maketitle
        \begin{abstract}
            We report an experimental study of the drying-induced peeling of a bilayer, consisting of an elastomeric disk coated with a suspension of nanoparticles.
            We show that although capillary forces associated with the scale of the droplet can not compete with the adhesion of the elastomer on a surface, nevertheless large tensile stresses develop in the coating, which results in a moment bending the bilayer.
            We attribute this stress to the nano-menisci in the pores of the colloidal material and we propose a model that describes successfully the early stage curvature of the bilayer.
            Thus, we show that the peeling can be conveniently controlled by the particle size and the coating thickness.
        \end{abstract}
    \end{@twocolumnfalse}
]

Many materials systems involve thin films or patches that cover substrate for the purposes of protection, functionalization, or hierarchical design.
In such cases, it is important to be able to overcome the adhesion of the film to substrate, as failure to do so can compromise an adhesion, damage material, etc.
In this paper, we highlight the use of a nanoparticle suspension, which, upon drying, produces stress and bending that assist the process of deadhesion.

To deform flexible slender bodies  {or thin films}, two main routes have been explored.
To estimate residual or cracking stresses of consolidating thin polymer films \cite{Thomas2011} or drying colloidal suspensions \cite{Yow2010}, the bending of a cantilever has been used \cite{Payne1998}.
Also, locally stressed materials can fold a thin film into a microstructured container \cite{Leong2008}, which has potential applications in microelectronics or biomaterial encapsulation.
Moreover, various stimuli have been used to induce  motion of slender structures with the use of solvent responsive, thermoresponsive or photoactive materials \cite{Holmes2011,Meng2013}.

Capillary forces exerted by a liquid drop can also bend slender elastic structures such as fibers \cite{Bico2004,Duprat2012} or sheets \cite{Huang2007,Py2007,Neukirch2013,Fargette2014}.
Considering a drop of surface tension  $\gamma$ sitting on a plate of thickness $h_s$ and elastic modulus $E$, we can define an elasto-capillary length \cite{Bico2004} $\ell_{EC} = \sqrt{Eh_s^3/\gamma}$, which sets the critical length of the sheet above which it will wrap around the drop.
This approach of capillary origami allows various self-folded polyhedra to be made with elastomers \cite{Py2007} or metals \cite{Leong2007}.

However, to successfully achieve capillary origami, it is important to make sure that either the adhesion of the elastomer with the surface is low or to force the de-adhesion by an impact of the drop on the elastomer \cite{Antkowiak2011}.
If adhesion forces to a substrate are larger than capillary forces, then it is impossible with a simple fluid to bend the slender elastic material.
Such a threshold is well known in peeling problems \cite{Kendall2004,Hamm2008}.

\begin{figure}
    \centering
    \includegraphics[scale=1]{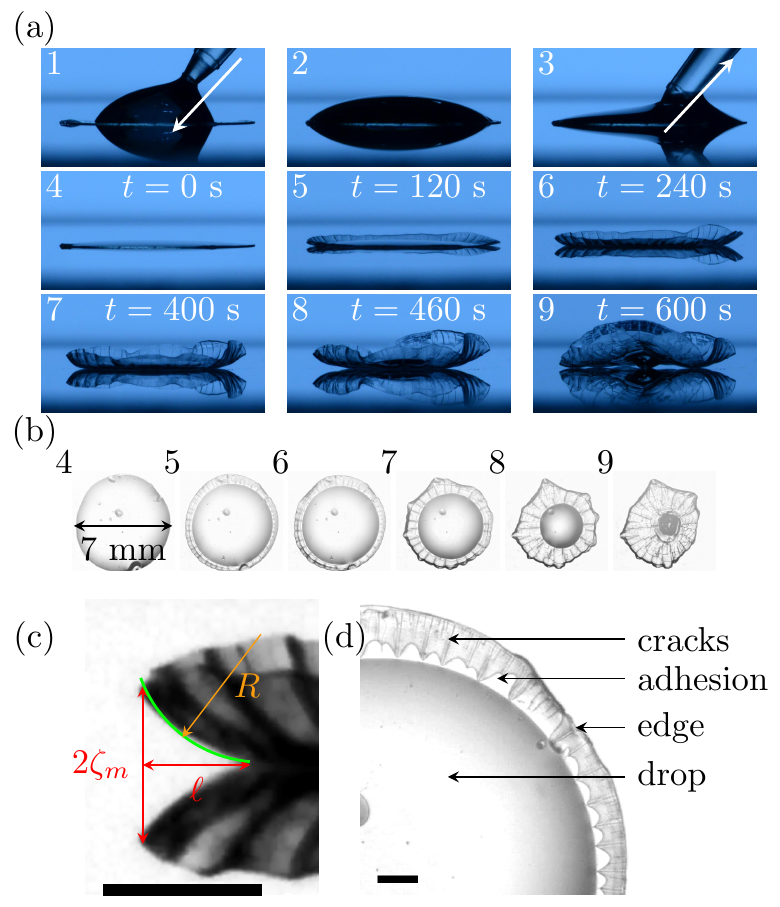}
    \caption{Illustration of a typical experiment with an elastomeric disk of $7$ mm in diameter and $16$ nm particles (HS) at $\phi_i=0.15$. (a) Side view. Pictures 1 to 3 show, respectively, the deposition of the colloidal droplet, the absence of debonding by capillary forces and the removal of a fraction of liquid to make a thin film. Pictures 4 to 6 illustrate the temporal evolution and the corresponding pictures in (b) are the bottom view. At $t=240$ s, enlargement of the curved bilayer (c) from a side view with a typical radius of curvature $R = \zeta_m^2/(2\ell)$ and (d) from the bottom view. The scale bar represents $0.5$ mm.}\label{fig:pictures}
    \end{figure}

    In this Letter, we investigate the bending of elastomeric sheets induced by stresses developed during the drying of colloidal suspensions.
    We show that where the usual elasto-capillary effect is unable to bend an elastic sheet because of the adhesion with a substrate, debonding becomes possible by the use of drying colloidal suspensions, which introduces the low pressure from many small menisci in pores and leads to large tensile stresses.

    \paragraph{Experimental protocol}

    The elastomers are prepared from polydimethylsiloxane (PDMS, Sylgard 184, Dow Corning) with a cross-linker/oligomer weight ratio of 1:10.
    The PDMS is spin-coated on the top of a polystyrene petri dish lid and cured at $70^\circ$C for $4$ hours and used within $3$ days.
    The resulting elastomer is then peeled of the lid and carefully transferred onto a rubber sheet with a particular attention to not stretch the film.
    The elastic modulus is $E_s=1.2$ MPa and film thicknesses have been measured by a white light spectrometer (OceanOptics USB2000+ used with a LS-1-LL tungsten halogen light source), confirmed by optical profilometry (Leica DCM3D) and range from $[50,125]$ \si{\micro\meter}.
    Punches are used to cut a disk in the film with a diameter of $7$ mm (Biopsy punches, Miltex).

    The elastomeric disk is then coated with a colloidal suspension (Fig. \ref{fig:pictures}(a)).
    In order to reproducibly make a flat coating, we follow several steps.
    The top surface of the disk is made hydrophilic with a plasma treatment and the elastomer is transferred to a transparent acrylic slide (McMaster-Carr, $4.5$ mm thickness).
    A drop of specified volume of colloidal suspension is deposited with a micropipette, e.g. $20$ \si{\micro\liter} for a $7$ mm disk diameter.
    This volume is chosen so that the drop easily spreads over the entire surface of the disk.
    A given excess of liquid is removed with a second micropipette, e.g. $16$ \si{\micro\liter}.

    The system is placed under an inverted microscope (Leica, $1.25\times$) and pictures are recorded with a camera (Edmund Optics, CMOS monochrome USB).
    A second camera (Nikon, Macro lens $105$ mm) records the time evolution of the shape from a side view.
    Observations are made at relative humidity of $50\pm5$\% and temperature $23^\circ$C.

    The colloidal suspensions we used are silica nanoparticles with different particle diameters $2a$ and their properties are summarized in Table \ref{tab:particles}.
    These suspensions can be diluted with deionized water at pH $9.5$, which is adjusted by the addition of NaOH.

    \begin{table}
        \begin{center}
            \begin{tabular}{|c|c|c|c|c|}
                \hline
                Name & $2a$ (nm) & $\phi_i$ & pH  \\
                \hline
                SM30 & 10 & 0.15  & 9.9 \\
                HS40 & 16 & 0.22 &  9.8  \\
                TM50 & 26 & 0.30 &  9.2 \\
                Kleb & 50 & 0.30  & 9.4 \\
                Lev & 92 & 0.30  & 8.9 \\
                \hline
            \end{tabular}
        \end{center}
        \caption{Properties of the colloidal suspensions of silica nanoparticles.
        The diameter is denoted $2a$ and the initial volume fraction $\phi_i$. SM30, HS40, TM50 are Ludox suspensions (purchased from Sigma-Aldrich), and Kleb and Lev denote Klebosol 50R50 and Levasil30, respectively.}\label{tab:particles}
    \end{table}

    \paragraph{Observations}

\begin{figure}
        \centering
        \includegraphics[scale=0.85]{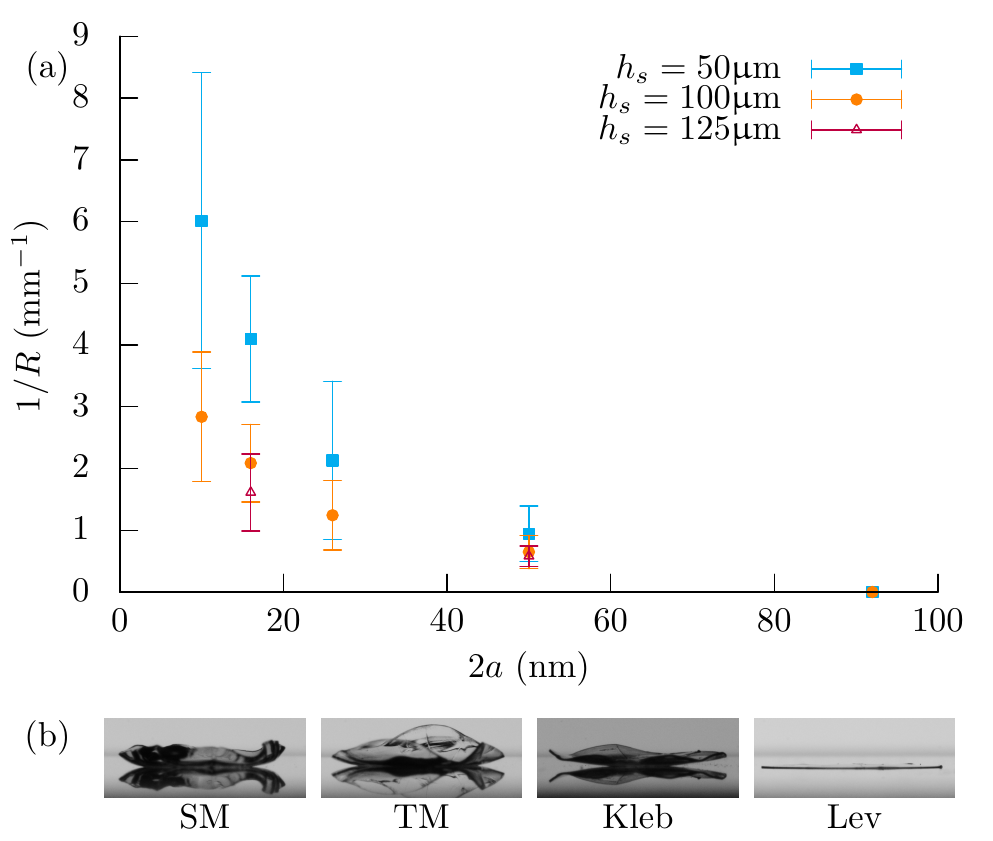} \\
        \caption{(a) Curvature $R^{-1}$ versus the particle diameter $2a$ of colloidal suspensions at an initial volume fraction  $\phi_i=0.15$ for different elastomer thicknesses $h_s$. The disk diameter is $7$ mm.
        (b) Final states (disk thickness: $50$ \si{\micro\meter}) for different colloidal suspensions at $\phi_i=0.15$.}\label{fig:curvature_raw}
    \end{figure}

\begin{figure}
        \centering
        \includegraphics[width=\columnwidth]{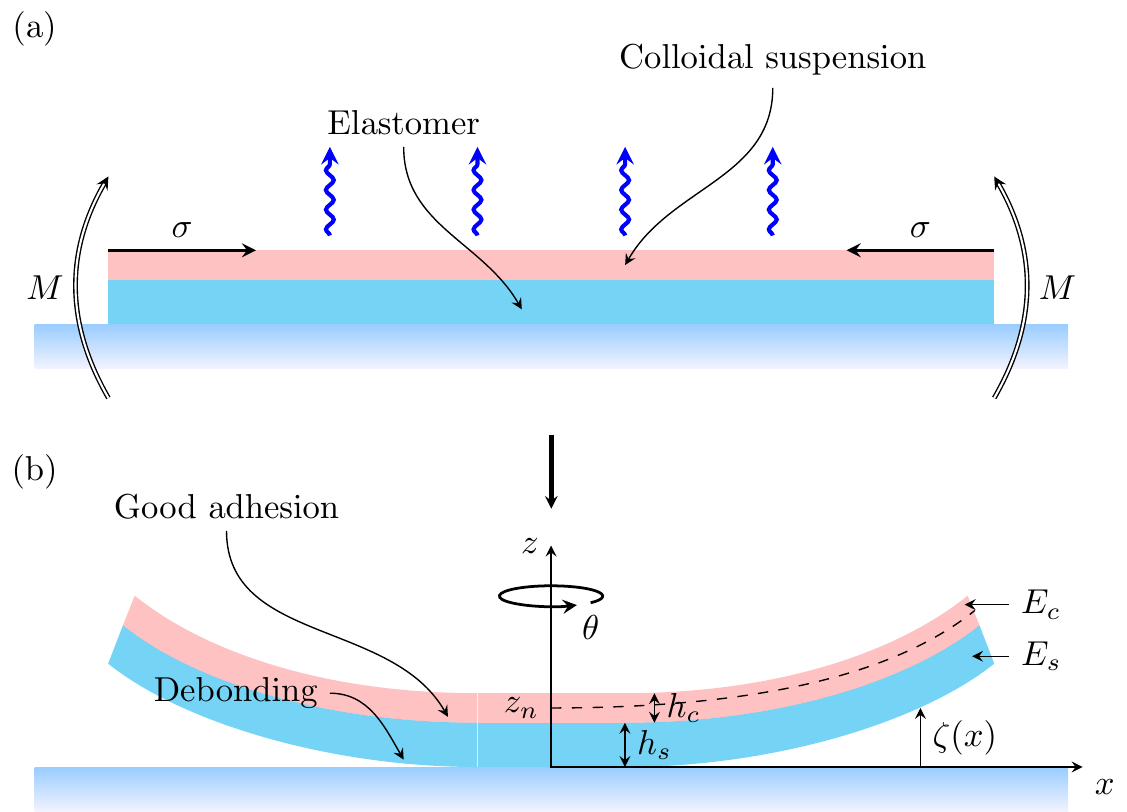}
        \caption{Sketch of the bilayer. (a) The drying of the colloidal suspension coated on the top of an elastomer induces tensile stresses.
            (b) The drying results in a moment $M$ peeling the bilayer which remains cohesive due to a good adhesion between colloids and the elastomer.
        }\label{fig:sketch}
    \end{figure}

    After the deposition of the colloidal suspension on the top of the PDMS disk, the liquid film dries from the edge, where the coating is thinner, to the center (Fig. \ref{fig:pictures}(b)) and exhibits radial crack patterns as is commonly seen in drying colloidal droplets \cite{Giorgiutti-Dauphine2014}.
    As the drying proceeds, tensile stresses develop in the coating, resulting in a moment on the bilayer material (Fig. \ref{fig:sketch}), which is peeled off the substrate with a  {macroscopic} radius of curvature $R$ as illustrated in Figs. \ref{fig:pictures}(a) and (c).
    This phenomenon is ensured by good adhesion of the colloidal material with the elastomer \cite{Leger1999}, which holds the two materials constituting the bilayer together.
    During the delamination, in front of the drying front, the contact area remaining between the elastomer and the surface adopts a pattern similar to that of a flower's petals (Fig. \ref{fig:pictures}(d)), which can be attributed to the cracks resulting in mechanical weaknesses in the bilayer.
    Thus, each coating strip can also bend the bilayer in the $\theta$ direction (defined in Fig. \ref{fig:sketch}) \cite{Zhuk1998}.
    Furthermore, the continuous peeling of the bilayer compresses its edge, which triggers a wrinkling instability \cite{Sharon2002,Aharoni2010} with a preferential location at the position of cracks in the coating.
    Figure \ref{fig:curvature_raw}(b) shows the final state with different colloidal particle sizes.
        Small particles induce a strong deformation of the elastomeric sheet that could lead to several small contact points whereas larger particles such as Klebosol have a weaker effect and form a single contact point in the center.
            Finally, above a critical particle size, the delamination is prevented and the bilayer remains flat.

  The P\'{e}clet number comparing advective and diffusive effects of the particles in the drying colloidal film can be written as $\textrm{Pe} = 6\pi \eta_s a h_c V_E/(k_B T)$, where $\eta_s$ is the solvent viscosity, $V_E$ the evaporation rate and $k_B T$ the thermal energy \cite{Routh1998}.
     {The drying rate of our colloidal suspension is independent of the particle size and has a typical measured value \cite{Boulogne2012a} of $V_E \sim 10^{-8}$ m/s. This estimate is confirmed by a scaling law $V_E \sim h_c/\tau \sim 10^{-8}$ m/s where $h_c\sim 10$ \si{\micro\meter} and $\tau\sim100$ s is the drying timescale (Fig. \ref{fig:pictures}(a)).
    Considering this drying rate, we have $\textrm{Pe}\sim 10^{-2}$, which means that we can assume that the particle distribution is homogeneous over the layer thickness.}

      Next, we focus on the peeling of the bilayer and we study its efficiency through the resulting early-stage curvature for different coatings and elastomer thicknesses.
        {
           In Fig. \ref{fig:curvature_raw}(a), we present measurements of the macroscopic curvature  $R^{-1}$ of the bilayer as presented in Fig. \ref{fig:pictures}(a).
           Measurements are performed for a length of the debonding front $\ell=[0.3,0.8]$ mm and we varied the particle sizes as well as the elastomer thickness.
   	To rationalize these results, we propose in the next paragraph a model relating the curvature of the bilayer and the drying stress.
    }

    \paragraph{Model}

    In a classical work, Stoney studied the stress induced by a coating on a deformable metal strip \cite{Stoney1909}, assuming that both layers have the same elastic modulus and that the coating thickness is much less than the substrate thickness.
    Various extensions of this work have been reported including the stress relief in the coating by bending \cite{Corcoran1969} and the shift of the position of the neutral axis due to different elastic moduli of the layers \cite{Chiu1990}.
    In our system, the elastomer is thicker and has a lower elastic modulus than the coating.
    Thus, we derive the equations under these assumptions \cite{Benabdi1997}.

    We assume that all quantities are independent of $\theta$ and consider the analogous problem of a one-dimensional model for a strip.
    The deflection of the material $\zeta(x)$ can be written via a Taylor series as:

    $$\zeta(x) = \zeta(x=0) + \partial_x \zeta(x=0) x + 1/2 \partial_x^2 \zeta(x=0) x^2 + ... $$
    The boundary conditions are that the material is clamped at $x=0$, which leads to $\zeta(x=0) =0$ and $\partial_x \zeta(x=0) = 0 $.
    Thus, we can deduce that the beam can be considered as a portion of a circle with a radius of curvature $R$:

    \begin{equation}
        \zeta(x) = \frac{1}{2} \frac{\partial^2 \zeta(x=0)}{\partial x^2} x^2 = \frac{x^2}{2R}.
    \end{equation}

    Denoting $\sigma$ the stress distribution in the beam, and $z$ the distance across th thin film, equilibrium of the sum of axial forces and  bending moments are, respectively, expressed by:

    \begin{eqnarray}
        \int \sigma \,\textrm{d}S = 0\label{eq:sum_force}\\
        \int \sigma  z\, \textrm{d}S = 0.\label{eq:sum_momentum}
    \end{eqnarray}

    Also, we denote the elementary cross-sectional area $\textrm{d}S = b \,\textrm{d}z$, where $b$ is a transverse dimension, $z_n$ the position of the neutral axis, $\sigma_c$ the stress developed in the coating by the drying, and $h_s$ and $h_c$ the thicknesses of the substrate and the elastomer, respectively (see Fig. \ref{fig:sketch}).
    From equation (\ref{eq:sum_force}), we have

    \begin{multline}
        \frac{E_s^\star b}{R} \int_{0}^{h_s} (z_n - z) \,\textrm{d}z
        + b \int_{h_s}^{h_s+h_c} \sigma_c \,\textrm{d}z \\
        + \frac{E_c^\star b}{R} \int_{h_s}^{h_s+h_c} (z_n - z) \,\textrm{d}z
        =0,\label{eq:integral}
    \end{multline}
    where $(z_n-z)/R$ is the strain and $E_i^\star = E_i/(1-\nu_i^2)$ with $\nu_i$ the Poisson ratio of the layer $i$.
    The second term accounts for the stress induced by the drying of the coating and the first and last terms represent the response stress of each bent layer.
    Assuming that $\sigma_c$ is independent of $z$, the integration of equation (\ref{eq:integral}) determines the position of the neutral axis $z_n$:

    \begin{equation}
        z_n - h_s =- \frac{\sigma_c R h_c}{E_c^\star h_c + E_s^\star h_s}
        + \frac{1}{2} \frac{E_s^\star h_s^2 - E_c^\star h_c^2}{E_c^\star h_c + E_s^\star h_s}.\label{eq:neutral_axis}
    \end{equation}

    Similarly, the integration of equation (\ref{eq:sum_momentum}) with the consideration of equation (\ref{eq:neutral_axis}) leads to:

    \begin{equation}
        \sigma_c = -\frac{E_s^\star h_s }{6 R} P (\Sigma, \Theta),\label{eq:stress}
    \end{equation}
where we defined the dimensionless numbers $\Sigma = E_c^\star/E_s^\star$ and $\Theta = h_c/h_s$, and the function

\begin{equation}
    P(\Sigma, \Theta) = \frac{ \Theta^{-1} + 2\Sigma( 2+3\Theta + 2\Theta^2 ) + \Sigma^2\Theta^3}{1+\Theta}.
\end{equation}.

    \begin{figure}
        \centering
        \includegraphics[scale=0.85]{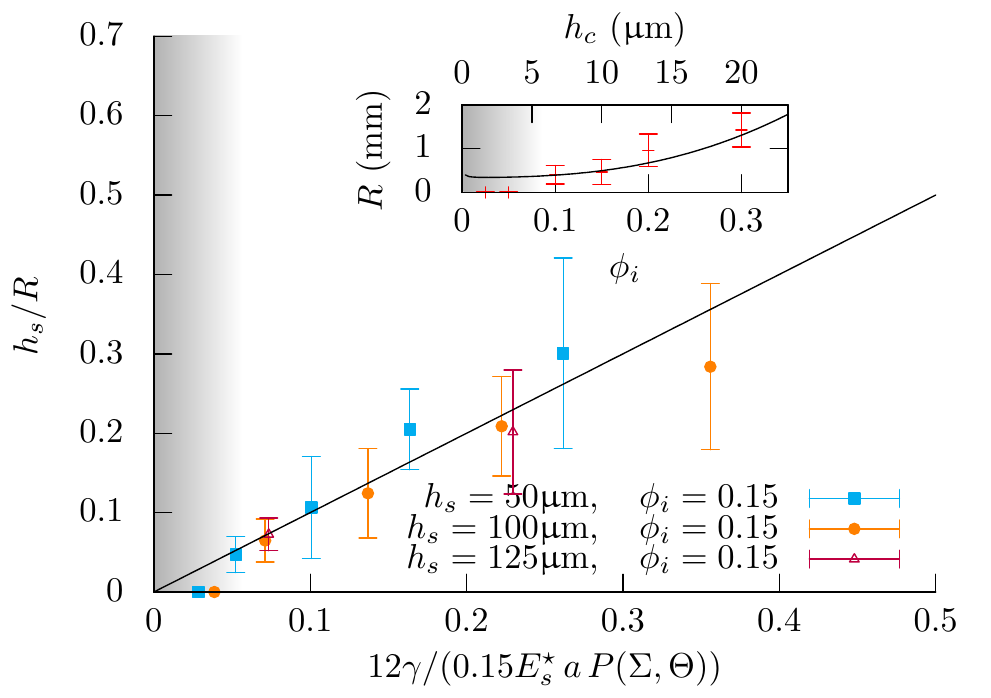} \\
        \caption{Dimensionless curvature ($h_s/R$) versus $12 \gamma / (0.15\,E_s^\star\,a P(\Sigma, \Theta))$ for different disk thicknesses $h_s$.
            The black solid lines represent equation (\ref{eq:curvature}) for $E_c=0.3$ GPa, without any additional prefactor.
            The inset represents the radius of curvature $R$ as a function of the initial volume fraction $\phi_i$ for $2a=26$ nm particles on a $7$ mm elastomer ($h_s=50$ \si{\micro\meter}).
            The shaded areas show the domain in which the peeling does not happen.
        }\label{fig:curvature}
    \end{figure}

    In the drying experiments, the air-water interface recedes into the film.
    We can estimate the stress produced by the nano-menisci between colloidal particles \cite{Routh1998,Dufresne2003}

    \begin{equation}
        \sigma_c \simeq -\frac{2 \gamma}{R_M} \simeq - \frac{2 \gamma}{0.15 a},\label{eq:stressmenisci}
    \end{equation}
    where $\gamma$ is the air-water surface tension, $R_M\simeq 0.15 a$ the maximum radius of the meniscus and $a$ the particle radius.

    Therefore, equation (\ref{eq:stressmenisci}) becomes
    \begin{equation}
        \frac{2 \gamma}{0.15 a}  \simeq \frac{E_s^\star h_s }{6 R} P(\Sigma,\Theta), \label{eq:curvature}
    \end{equation}
    which means that the radius of curvature $R\propto a$.
    Thus, the particle size is a convenient control variable.

    This prediction is compared to experimental observations on acrylic slides for different elastomer thicknesses $h_s$, coating thicknesses $h_c$ and particle radii $a$.
    Note that the coating thickness depends on the initial volume fraction of the colloidal suspension  {and also on the radial position}.

     {
    To quantify the effect of the coating thickness, we prepared the elastomeric disk in the same manner as presented in the paper.
These disks are glued on glass slides by a plasma treatment to form chemical bonds.
To make the disk surface hydrophilic, a second plasma treatment is used.
However, to prevent the glass slide being hydrophilic, the glass slide not covered by the disk is protected with tape.
This step is necessary to have similar conditions to the main experiment and to avoid the liquid to wet easily the glass slide.
Thus, it is still possible to make a large drop as shown in step 2 of Fig. \ref{fig:pictures}(a).
After complete drying, we measured the shape of the coating near the disk edge with an optical profiler.
Thus, an average value of the coating thickness over the length of the debonding front $\ell=[0.3,0.8]$ mm is deduced with an uncertainty of $3$ \si{\micro\meter}.
We relate this thickness to the relevant coating thickness $h_c$ in the delamination problem because Fig. 1(d) shows that the peeling occurs when the material is at an advanced step of the drying process.
 The coating thickness is independent of the particle size and for $\phi_i=0.15$, it is about $10$ \si{\micro\meter}.
}

    With this approach, in Fig. \ref{fig:curvature}(a), we report measurements of the radius of curvature averaged for different experiments as a function of $a$.
     {We must note that the measured radius of curvature is macroscopic. As shown in figure \ref{fig:pictures}(c), some deviations are observed, that can be attributed to the inhomogeneous coating thickness.}

             {The value of the elastic modulus of drying silica suspension is not well established in the literature \cite{Goehring2013} due to several experimental difficulties: heterogeneities of the material, temporal evolution due to drying, possible effects of the drying conditions such as the geometry or the relative humidity.
            We assume that $E_c$ is a weak function of the particle size $a$ in the studied range.}
            We observe that {our experimental results (Fig. \ref{fig:curvature})} are well described by the theoretical prediction given by equation (\ref{eq:curvature}) with the fitted parameter $E_c=0.3$ GPa (with $\nu_s=0.5$ and $\nu_c=0.3$) in agreement with values available in the literature \cite{Zarzycki1988,Boulogne2012a}.
The collapse of our data onto a single curve is consistent with the weak effect of the particle size on the elastic modulus.

Moreover, we notice the limit for which the adhesion prevents the peeling of the bilayer (Fig. \ref{fig:curvature_raw}(b) and \ref{fig:curvature}).
    This is encountered either for large particles, which induce smaller drying tensile stresses (equation (\ref{eq:stressmenisci})), or by a thin coating, as illustrated by equation (\ref{eq:stress}) and the inset of Fig. \ref{fig:curvature}.

    \paragraph{Conclusion}

    \begin{figure}
        \centering
        \includegraphics[scale=1]{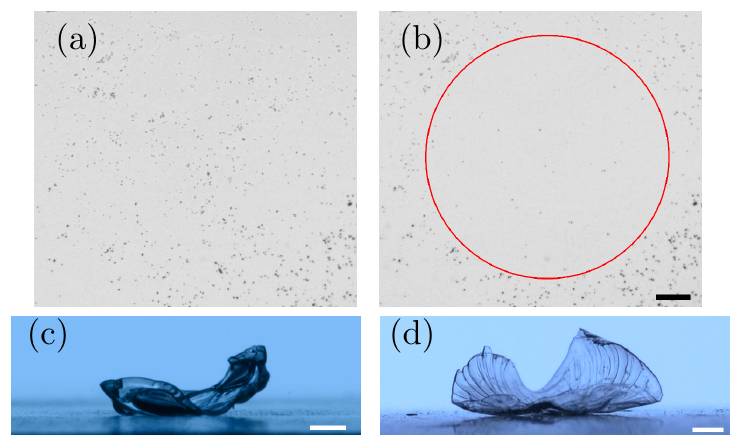}
        \caption{Illustrations of the technique on various surfaces with a PDMS disk with a diameter of $7$ mm and a thickness of $50$ \si{\micro\meter} coated with a HS suspension at $\phi_i=0.15$.
            (a) Dust is randomly distributed on an acrylic slide.
            (b) After the cleaning of the surface, more than $90$\% of the particles have been removed.
            The red circle shows the initial position of the elastomeric disk.
            (c) Final state on a printed paper and
        (d) on varnished wood (oak). Scale bars represent $1$ mm.}\label{fig:applications}
    \end{figure}


     {For a droplet deposited on the top of a thin elastomeric sheet, we have shown that capillary forces are not sufficient to compete with adhesion and  capillary origami does not occur.
    Thus, in such situation for which it is interesting to remove the sheet from the substrate, we suggest that colloidal nanoparticles can be used to induce large tensile stresses that are able to peel the elastomer from the substrate.}

    We provide a model predicting the curvature of the bilayer, which is in a good agreement with experimental results.
    In particular, the model shows the effect of the coating thickness as well as the particle size is responsible for the tensile stress.
    As a result, drying tensile stresses higher than capillary stresses can be achieved with the use of nanoparticles.

    We believe that these colloidal systems can be employed in various situations where delicate surfaces are involved such as in applications with optical and electronic components or in restoration of photographs, painting, wallpaper and fragile collectibles from contamination by dust, pollen, dirt, etc. \cite{Carretti2005,Carretti2010}.
    For illustrative purposes, we show in Fig. \ref{fig:applications} that the bilayer studied in this Letter can be employed to remove silica particles (typical size: $50$ \si{\micro\meter}) randomly distributed on an acrylic slide with a success larger than $90$\% in a single step.  {The advantage of the method relies on the absence of an external intervention to remove the material, which decreases the possibility to damage the surface.}
    We also show that the method works on ink-printed paper or a varnished wood surface, which represent potential applications for ancient manuscripts or antique furnitures.


    \paragraph{Acknowledgements}
    We are very grateful to F. Giorgiutti-Dauphin\'{e} and L. Pauchard for enlightening discussions about delamination and their applications. We also thank J.-M. Copin and M. Persson for providing Klebosol and Levasil samples respectively.
    F.B. acknowledges that the research leading to these results received funding from the People Programme (Marie Curie Actions) of the European Union's Seventh Framework Programme (FP7/2007-2013) under REA grant agreement 623541.
    We thank the NSF for partial support via grant CBET-1132835.

    \bibliography{article}

\begin{thebibliography}{10}

\bibitem{Thomas2011}
K.R. Thomas and U.~Steiner.
\newblock {D}irect stress measurements in thin polymer films.
\newblock {\em Soft Matter}, 7:7839--7842, 2011.

\bibitem{Yow2010}
H.N. Yow, M.~Goikoetxea, L.~Goehring, and A.F. Routh.
\newblock {E}ffect of film thickness and particle size on cracking stresses in
  drying latex films.
\newblock {\em J. Coll. and Int. Sci.}, 352:542--548, 2010.

\bibitem{Payne1998}
J.A. Payne.
\newblock {\em {S}tress {E}volution in {S}olidifying {C}oatings}.
\newblock PhD thesis, 1998.

\bibitem{Leong2008}
T.G. Leong, B.R. Benson, E.K. Call, and D.H. Gracias.
\newblock {T}hin film stress driven self-folding of microstructured containers.
\newblock {\em Small}, 4:1605--1609, 2008.

\bibitem{Holmes2011}
D.P. Holmes, M.~Roche, and T.~Sinha.
\newblock {B}ending and twisting of soft materials by non-homogenous swelling.
\newblock {\em Soft Matter}, 7:5188--5193, 2011.

\bibitem{Meng2013}
H.~Meng, H.~Mohamadian, M.~Stubblefield, D.~Jerro, S.~Ibekwe, S.-S. Pang, and
  G.~Li.
\newblock {V}arious shape memory effects of stimuli-responsive shape memory
  polymers.
\newblock {\em Smart Materials and Structures}, 22:093001, 2013.

\bibitem{Bico2004}
J.~Bico, B.~Roman, L.~Moulin, and A.~Boudaoud.
\newblock {A}dhesion: {E}lastocapillary coalescence in wet hair.
\newblock {\em Nature}, 432:690--690, 2004.

\bibitem{Duprat2012}
C.~Duprat, S.~Protiere, A.Y. Beebe, and H.A. Stone.
\newblock {W}etting of flexible fibre arrays.
\newblock {\em Nature}, 482:510--513, 2012.

\bibitem{Huang2007}
J.~Huang, M.~Juszkiewicz, W.~De~Jeu, E.~Cerda, T.~Emrick, N.~Menon, and
  T.~Russell.
\newblock {C}apillary wrinkling of floating thin polymer films.
\newblock {\em Science}, 317:650--653, 2007.

\bibitem{Py2007}
C.~Py, P.~Reverdy, L.~Doppler, J.~Bico, B.~Roman, and C.N. Baroud.
\newblock {C}apillary origami: spontaneous wrapping of a droplet with an
  elastic sheet.
\newblock {\em Phys. Rev. Lett.}, 98:156103, 2007.

\bibitem{Neukirch2013}
S.~Neukirch, A.~Antkowiak, and J.-J. Marigo.
\newblock {T}he bending of an elastic beam by a liquid drop: a variational
  approach.
\newblock {\em Proceedings of the Royal Society A}, 469:20130066, 2013.

\bibitem{Fargette2014}
A.~Fargette, S.~Neukirch, and A.~Antkowiak.
\newblock {E}lastocapillary snapping: capillarity induces snap-through
  instabilities in small elastic beam.
\newblock {\em Phys. Rev. Lett.}, 112:137802, 2014.

\bibitem{Leong2007}
T.~Leong, P.~Lester, T.~Koh, E.~Call, and D.~Gracias.
\newblock {S}urface tension-driven self-folding polyhedra.
\newblock {\em Langmuir}, 23:8747--8751, 2007.

\bibitem{Antkowiak2011}
A.~Antkowiak, B.~Audoly, C.~Josserand, S.~Neukirch, and M.~Rivetti.
\newblock {I}nstant fabrication and selection of folded structures using drop
  impact.
\newblock {\em Proc. of the Nat. Acad. of Sci.}, 108:10400--10404, 2011.

\bibitem{Kendall2004}
K.~Kendall.
\newblock {\em {F}ilms and {L}ayers: {A}dhesion of {C}oatings}, pages 327--351.
\newblock Springer, 2004.

\bibitem{Hamm2008}
E.~Hamm, P.~Reis, M.~LeBlanc, B.~Roman, and E.~Cerda.
\newblock {T}earing as a test for mechanical characterization of thin adhesive
  films.
\newblock {\em Nature Mater.}, 7:386--390, 2008.

\bibitem{Giorgiutti-Dauphine2014}
F.~Giorgiutti-Dauphin{\'e} and L.~Pauchard.
\newblock {E}lapsed time for crack formation during drying.
\newblock {\em The European Physical Journal E}, 37:1--7, 2014.

\bibitem{Leger1999}
L.~L{\'e}ger, E.~Rapha{\"e}l, and H.~Hervet.
\newblock {\em {S}urface-anchored polymer chains: their role in adhesion and
  friction}, pages 185--225.
\newblock Springer Berlin Heidelberg, 1999.

\bibitem{Zhuk1998}
A.V. Zhuk, A.G. Evans, J.W. Hutchinson, and G.M. Whitesides.
\newblock {T}he adhesion energy between polymer thin films and self-assembled
  monolayers.
\newblock {\em Journal of Materials Research}, 13:3555--3564, 1998.

\bibitem{Sharon2002}
E.~Sharon, B.~Roman, M.~Marder, G.-S. Shin, and H.L. Swinney.
\newblock {B}uckling cascades in free sheets.
\newblock {\em Nature}, 419:579--579, 2002.

\bibitem{Aharoni2010}
H.~Aharoni and E.~Sharon.
\newblock {D}irect observation of the temporal and spatial dynamics during
  crumpling.
\newblock {\em Nature Mater.}, 9:993--997, 2010.

\bibitem{Routh1998}
A.F. Routh and W.B. Russel.
\newblock {H}orizontal drying fronts during solvent evaporation from latex
  films.
\newblock {\em AIChE J.}, 44:2088--2098, 1998.

\bibitem{Boulogne2012a}
F.~Boulogne, L.~Pauchard, and F.~Giorgiutti-Dauphin{\'e}.
\newblock {E}ffect of a non-volatile cosolvent on crack patterns induced by
  desiccation of a colloidal gel.
\newblock {\em Soft Matter}, 8:8505--8510, 2012.

\bibitem{Stoney1909}
G.G. Stoney.
\newblock {T}he tension of metallic films deposited by electrolysis.
\newblock {\em Proceedings of the Royal Society of London. Series A},
  82:172--175, 1909.

\bibitem{Corcoran1969}
E.~M. Corcoran.
\newblock {D}etermining stresses in organic coatings using plate beam
  deflection.
\newblock {\em J. of Paint Tech.}, 41:635, 1969.

\bibitem{Chiu1990}
C.-C. Chiu.
\newblock {D}etermination of the elastic modulus and residual stresses in
  ceramic coatings using a strain gage.
\newblock {\em J. Am. Ceram. Soc.}, 73:1999--2005, 1990.

\bibitem{Benabdi1997}
M.~Benabdi and A.A. Roche.
\newblock {M}echanical properties of thin and thick coatings applied to various
  substrates. {P}art {I}. {A}n elastic analysis of residual stresses within
  coating materials.
\newblock {\em Journal of Adhesion Science and Technology}, 11:281--299, 1997.

\bibitem{Dufresne2003}
E.R. Dufresne, E.I. Corwin, N.A. Greenblatt, J.~Ashmore, D.Y. Wang, A.D.
  Dinsmore, J.X. Cheng, X.S. Xie, J.W. Hutchinson, and D.A. Weitz.
\newblock {F}low and fracture in drying nanoparticle suspensions.
\newblock {\em Phys. Rev. Lett.}, 91:224501, 2003.

\bibitem{Goehring2013}
L.~Goehring, W.~Clegg, and A.~Routh.
\newblock {P}lasticity and fracture in drying colloidal films.
\newblock {\em Phys. Rev. Lett.}, 110:024301, 2013.

\bibitem{Zarzycki1988}
J.~Zarzycki.
\newblock {C}ritical stress intensity factors of wet gels.
\newblock {\em J. Non-Crystalline Solids}, 100:359--363, 1988.

\bibitem{Carretti2005}
E.~Carretti, L.~Dei, and R.G. Weiss.
\newblock {S}oft matter and art conservation. {R}heoreversible gels and beyond.
\newblock {\em Soft Matter}, 1:17--22, 2005.

\bibitem{Carretti2010}
E.~Carretti, M.~Bonini, L.~Dei, B.H. Berrie, L.V. Angelova, P.~Baglioni, and
  R.G. Weiss.
\newblock {N}ew frontiers in materials science for art conservation: responsive
  gels and beyond.
\newblock {\em Accounts of Chemical Research}, 43:751--760, 2010.

\end{thebibliography}
    \bibliographystyle{unsrt}

    \end{document}